\documentclass[preprint2]{aastex}

\usepackage{epstopdf}
\usepackage{amsmath}
\usepackage[tight]{subfigure}
\usepackage[utf8]{inputenc} 
\usepackage{graphicx}

\newcommand{\ang}{\AA~}
\defcitealias{barnes_inference_2016-1}{Paper II}

\begin{document}
	\title{Inference of Heating Properties from ``Hot'' Non-flaring Plasmas in Active Region Cores I. Single Nanoflares}
	\shorttitle{``Hot'' Non-flaring Plasmas I. Single Nanoflares}
	\author{W. T. Barnes\altaffilmark{1}}
	\author{P. J. Cargill\altaffilmark{2,3}}
	\and
	\author{S. J. Bradshaw\altaffilmark{1}}
	\altaffiltext{1}{Department of Physics \& Astronomy, Rice University, Houston, TX 77251-1892; will.t.barnes@rice.edu, stephen.bradshaw@rice.edu}
	\altaffiltext{2}{Space and Atmospheric Physics, The Blackett Laboratory, Imperial College, London SW7 2BW; p.cargill@imperial.ac.uk}
	\altaffiltext{3}{School of Mathematics and Statistics, University of St Andrews, St Andrews, Scotland KY16 9SS}
	\begin{abstract}
		The properties expected of ``hot'' non-flaring plasmas due to nanoflare heating in active regions are investigated using hydrodynamic modeling tools, including a two-fluid development of the EBTEL code. Here we study a single nanoflare and show that while simple models predict an emission measure distribution extending well above 10 MK that is consistent with cooling by thermal conduction, many other effects are likely to limit the existence and detectability of such plasmas. These include: differential heating between electrons and ions, ionization non-equilibrium and, for short nanoflares, the time taken for the coronal density to increase. The most useful temperature range to look for this plasma, often called the ``smoking gun'' of nanoflare heating, lies between $10^{6.6}$ and $10^7$ K. Signatures of the actual heating may be detectable in some instances.
	\end{abstract}
	\section{Introduction}
	\label{sec:intro}
	\par Observations of the magnetically closed solar corona from the \textit{Hinode} \citep{kosugi_hinode_2007} and Solar Dynamics Observatory (SDO) \citep{pesnell_solar_2012} spacecraft have led, for the first time, to quantitative studies of the distribution of coronal plasma as a function of temperature, and preliminary deductions about the heating process \citep[see papers in][]{de_moortel_recent_2015}. The key to this has been the ability to make measurements of the corona over a wide range of temperatures from the EUV Imaging Spectrometer (EIS) \citep{culhane_euv_2007} and X-Ray Telescope (XRT) \citep{golub_x-ray_2007} instruments on \textit{Hinode}, and the Atmospheric Imaging Assembly (AIA) \citep{lemen_atmospheric_2012} on SDO. Underpinning this work is the concept of nanoflare heating of the corona. Nanoflares \citep[e.g.][]{parker_nanoflares_1988} are small bursts of energy release, which, despite the implication in their name, have unknown magnitude and duration. While commonly attributed to small-scale magnetic reconnection, nanoflares can occur in other heating scenarios \citep[e.g.][]{ofman_self-consistent_1998}.
	\par One example of this approach has been studies of active region (AR) core loops \citep{warren_constraints_2011,warren_systematic_2012,winebarger_using_2011,tripathi_emission_2011,schmelz_cold_2012,bradshaw_diagnosing_2012,reep_diagnosing_2013,del_zanna_evolution_2015}. These are the brightest structures in ARs, spanning the magnetic polarity line, and are observed over a wide range of temperatures. An important result has been the determination of the emission measure distribution as a function of temperature ($\mathrm{EM}(T)\sim n^2dh$) along a line of sight. These workers showed that the emission measure peaked at $T = T_m = 10^{6.5}$ – $10^{6.6}$ K with $\mathrm{EM}(T_m)$ of order $10^{27}$ – $10^{28}$ cm$^{-5}$.  Below $T_m$ a relation of the form $\mathrm{EM} \propto T^a$ was found, with $2 < a < 5$. This distribution can be understood by a combination of radiative cooling of the corona to space and an enthalpy flux to the transition region (TR) \citep[e.g.][]{bradshaw_cooling_2010,bradshaw_new_2010} and and has significant implications for nanoflare heating. Defining low and high frequency (LF and HF) nanoflares by the ratio of the average time between nanoflares on a magnetic strand or sub-loop ($\langle t_N \rangle$) to the plasma cooling time from the peak emission measure ($\tau_{cool}$), LF (HF) nanoflares have $\langle t_N \rangle > (<) \tau_{cool}$ respectively. LF nanoflares have $a \sim$ 2 - 3 and thus do not account for many of the observations. In fact, \citet{cargill_active_2014} argued that these results implied a heating mechanism with $\langle t_N \rangle$ of order 1000 - 2000 s between nanoflares, with the value of $t_N$ associated with each nanoflare being proportional to its energy. Such intermediate frequency (IF) nanoflares have different energy build-up requirements from the commonly assumed LF scenario \citep{cargill_active_2014}.
	\par A second outcome of AR studies is the detection of a ``hot'' non-flaring coronal component characterised by plasma with $T > T_m$, a long-predicted consequence of nanoflare heating \citep{cargill_implications_1994,cargill_diagnostics_1995}. This has been identified from \textit{Hinode} and SDO data \citep{reale_evidence_2009,schmelz_hinode_2009,testa_hinode/eis_2012}, and retrospectively from data obtained by the X-Ray Polychrometer (XRP) instrument flown on the Solar Maximum Mission \citep{del_zanna_elemental_2014}. While characterising this emission is difficult \citep[e.g.][]{testa_temperature_2011,winebarger_defining_2012}, a similar scaling, $\mathrm{EM} \propto T^{-b}$ has been claimed \citep[e.g.][]{warren_systematic_2012}, with $b$ of order 7 – 10, though \citeauthor{del_zanna_elemental_2014} find larger values. \citeauthor{warren_systematic_2012} quote typical errors of $\pm$ 2.5 - 3 on these values due to the very limited data available above $T_m$ and \citeauthor{winebarger_defining_2012} have noted that the paucity of data from \textit{Hinode} at these temperatures could be missing significant quantities of plasma with $T > T_m$.
	\par In an effort to diminish uncertainty in this high temperature ``blind spot'' in $\mathrm{EM}(T)$, \citet{petralia_thermal_2014} analyzed an AR core by supplementing EIS spectral observations with broadband AIA and XRT measurements. By using concurrent observations from the 94 \ang channel of AIA and the Ti\_poly filter of XRT, the authors showed that the $\mathrm{EM}(T)$ peaked near $T_m = 10^{6.6}$ and had a weak, hot component. Additionally, \citet{miceli_x-ray_2012}, using the SphinX instrument \citep{sylwester_sphinx:_2008,gburek_sphinx_2011}, analyzed full-disk X-ray spectra integrated over 17 days, during which time two prominent ARs were present. These authors found that a two-temperature model was needed to fit the resulting spectrum, a strong 3 MK component and a much weaker 7 MK component.
	\par More recent data has come from rocket flights. The Focusing Optics X-ray Solar Imager (FOXSI) \citep{krucker_focusing_2013} first flew in November 2012 and observed an AR. A joint study with EIS and XRT by \citet{ishikawa_constraining_2014} suggested that while hot plasma existed up to 10 MK, the \textit{Hinode} instruments over-estimated the amount of plasma there. A rocket flight reported by \citet{brosius_pervasive_2014} identified emission in an Fe XIX line with peak formation temperature of $10^{6.95}$ K and reported an emission measure that was 0.59 times the emission formed at $10^{6.2}$ K. More recently, a pair of rocket flights gave observations from the Amptek X123-SDD soft X-ray spectrometer \citep{caspi_new_2015}. This provided comprehensive coverage of the 3 - 60 \ang wavelength range. \citeauthor{caspi_new_2015} demonstrated that the emission in this range could be fit by an emission measure with a power-law distribution slope of roughly $b = 6$. While all of these observations are very suggestive of nanoflare heating, it should also be noted that pixel-averaging, long time averages and/or inadequate instrument spatial resolution may lead to contamination of the $\mathrm{DEM}$ by multiple structures along the line of sight. It is desirable to obtain future measurements of plasma emission at $T>T_m$ from a single structure, such as a core active region loop, along the line of sight.
	\par Several other workers have combined model results with observations in an effort to better elucidate nanoflare signatures. Using a hydrodynamic loop model, \citet{reale_solar_2011} showed that emission from impulsively heated subarcsecond strands is finely structured and that this predicted structure can also be found in AR core emission as observed by the 94 \ang channel of AIA. Most recently, \citet{tajfirouze_time-resolved_2016}, using a 0D hydrodynamic model, explored a large parameter space in event energy distribution, pulse duration, and number of loops. Using a probabilistic neural network, the authors compared their many forward-modeled light curves to 94 \ang AIA observations of a ``hot'' AR core. They found that the observed light curves were most consistent with a pulse duration of 50 s and a shallow event energy distribution, suggestive of nanoflare heating.
	\par While the distributions of temperature and density above $T_m$ are likely to be determined by nanoflare heating and conductive cooling, there are several complications arising from the low density and high temperature present there. These are (i) the breakdown of the usual Spitzer description of thermal conduction which leads to slower conductive cooling, (ii) recognition that in cases of heating in a weakly collisional or collisionless plasma, electrons and ions need not have the same temperature since when one is heated preferentially the time for the temperature to equilibrate is longer than the electron conductive cooling time, and (iii) a lack of ionization equilibrium that can underestimate the quantity of the plasma with a given electron temperature.
	\par Thus the aim of the present and following paper, \citet[in preparation]{barnes_inference_2016-1} \citepalias[hereafter]{barnes_inference_2016-1}, is to investigate this high temperature regime from a modeling viewpoint with the aim of obtaining information that can be of use in the interpretation of present and future observations. In this paper we focus on single-nanoflare simulations and build up an understanding of the role of the different pieces of physics. \citetalias{barnes_inference_2016-1} addresses the properties of nanoflare trains. Given the limitations of present observations, the results of both papers are in part predictive for a future generation of instruments. \autoref{sec:phys_sum} addresses our methodology, including simple outlines of the physics expected from conductive cooling, the preferred heating of different species, and ionization non-equilibrium. \autoref{sec:results} shows results for our single- and two-fluid models, and \autoref{sec:discussion} provides discussion of the main points of our results.
	\section{Summary of Relevant Physics}
	\label{sec:phys_sum}
	\par We begin by considering the situation when a coronal loop (or sub-loop) cools in response to a nanoflare by the evolution of a single-fluid plasma $(T_e = T_i)$ along a magnetic field line. We deal with the case of electron-ion non-equilibrium in  \autoref{subsec:two_fluid_theory}. The energy equation is,
	\begin{equation}
		\label{eq:energy_1d}
		\frac{\partial E}{\partial t} = -\frac{\partial}{\partial s}[v(E+P)] - \frac{\partial F_c}{\partial s} + Q - n^2\Lambda(T),
	\end{equation}
where $v$ is the velocity, $E=p/(\gamma -1) + \rho v^2/2$, $F_c=-\kappa_0 T^{5/2}\partial T/\partial s$ is the heat flux, $Q$ is a heating function that includes both steady and time-dependent components, $\Lambda(T)=\chi T^{\alpha}$ is the radiative loss function in an optically thin plasma \citep[e.g.][]{klimchuk_highly_2008} and $s$ is a spatial coordinate along the magnetic field. In addition the equations of mass and momentum conservation are solved. These equations are closed by $p=2nk_BT$, the equation of state. For a given initial state and $Q$, the plasma evolution can then be followed.

\par In this paper, two approaches are used to solve \autoref{eq:energy_1d}. One uses the HYDRAD code \citep{bradshaw_influence_2013} which solves the full field-aligned hydrodynamic two-fluid equations. The second develops further the zero-dimensional Enthalpy Based Thermal Evolution of Loops (EBTEL) approach which solves for average coronal plasma quantities \citep{klimchuk_highly_2008,cargill_enthalpy-based_2012,cargill_enthalpy-based_2012-1,cargill_modelling_2015}. In this paper we compare the HYDRAD and EBTEL results and outline some restrictions that apply to the use of EBTEL when modeling the hot coronal component. However, the value of the EBTEL approach lies in its simplicity and computational speed, and the consequent ability to model the corona as a multiplicity of thin loops for long times, as we do in \citetalias{barnes_inference_2016-1}. Such calculations remain challenging for field-aligned hydrodynamic models.
	\par The derivation of the single-fluid EBTEL equations can be found in \citep{klimchuk_highly_2008,cargill_enthalpy-based_2012}. We assume subsonic flows, and \autoref{eq:energy_1d} and the equation of mass conservation are solved for nanoflare energy input. EBTEL treats the corona and TR as separate regions, matched at the top of the TR by continuity of conductive and enthalpy fluxes. It produces spatially-averaged, time-dependent quantities (e.g. $\bar{T}(t),\bar{n}(t)$) in the corona and can also compute quantities at the loop apex and the corona/TR boundary. The single-fluid EBTEL equations are,
	\begin{align}
		\frac{1}{\gamma - 1}\frac{d\bar{p}}{dt} =& \,\bar{Q} - \frac{1}{L}(\mathcal{R}_C + \mathcal{R}_{TR}), \label{eq:energy_0d} \\
		\frac{\gamma}{\gamma - 1}(pv&)_0 + F_{c,0} + \mathcal{R}_{TR} = 0, \label{eq:tr_energy_0d} \\
		\frac{d\bar{n}}{dt} =& -\frac{c_2(\gamma - 1)}{2c_3\gamma Lk_B\bar{T}}(F_{c,0} + \mathcal{R}_{TR}).\label{eq:mass_0d}
	\end{align}
Here an overbar denotes a coronal average, $F_{c,0} = -(2/7)\kappa_0 T_a^{7/2}/L$ is the heat flux at the top of the TR (see also \autoref{subsec:hf_theory}), $\mathcal{R}_C=\bar{n}^2\Lambda(\bar{T})L$, is the integrated coronal radiation, $\mathcal{R}_{TR}$ is the integrated TR radiation, and $L$ is the loop half-length. The subscript ``0'' denotes a quantity at the top of the TR and ``$a$'' denotes a quantity at the loop apex. Solving this set of equations requires the specification of three (semi-)constants that are defined by  $c_1=\mathcal{R}_{TR}/\mathcal{R}_C$, $c_2=\bar{T}/T_a$ and $c_3=T_0/T_a$. $c_2$ and $c_3$ can be taken as constant, with values of 0.9 and 0.6 respectively. \citet{cargill_enthalpy-based_2012} discuss the full implementation of $c_1 = c_1(T_a,L)$. \autoref{appendix_c1_corrections} provides a detailed discussion of the additional corrections we have applied to $c_1$ in order to ensure better agreement with HYDRAD for impulsive heating scenarios. \autoref{eq:energy_0d} is a statement of energy conservation in the combined corona and TR. \autoref{eq:tr_energy_0d} is the TR energy equation: if the heat flux into the TR is greater (smaller) than its ability to radiate then there is an enthalpy flux into (from) the corona. \autoref{eq:mass_0d} combines \autoref{eq:tr_energy_0d} with that of mass conservation.

	\subsection{Heat Flux Limiters}
	\label{subsec:hf_theory}
	\par It is well known that thermal conduction deviates from the familiar Spitzer-H{\"a}rm formula \citep{spitzer_transport_1953} at high temperatures \citep[e.g.][]{ljepojevic_heat_1989}. There is a firm upper limit on the heat flux: the free-streaming limit, $F_s=(1/2)fnk_BTV_e$, where $V_e$ is the electron thermal speed and $f$, a dimensionless constant, is determined from a combination of lab experiments, theory, and numerical models. The free-streaming flux is included in EBTEL and HYDRAD by a simple modification \citep{klimchuk_highly_2008},
	\begin{equation}
		F_{c,0} = \frac{F_cF_s}{\sqrt{F_c^2 + F_s^2}},
	\end{equation}
where $F_c$ is the Spitzer-H{\"a}rm heat flux. Smaller values of $f$ limit the heat flux to a greater degree. There is some disagreement on the optimal value of $f$. \citet{luciani_nonlocal_1983} use $f=0.1$ while \citet{karpen_nonlocal_1987} use $f=0.53$, and \citet{patsourakos_coronal_2005} choose $f=1/6$. Unless explicitly stated otherwise, we use $f = 1$ in order to compare EBTEL results with those of HYDRAD \citep[see appendix of][]{bradshaw_influence_2013}. The main aspect of inclusion of a free-streaming limit is to slow down conductive cooling. We do not consider here other conduction models \citep[e.g. the non-local model discussed in the coronal context by][]{karpen_nonlocal_1987,ciaravella_non-local_1991,west_lifetime_2008} since they lead to similar generic results.
	\subsection{Two-fluid Modeling}
	\label{subsec:two_fluid_theory}
	\par In some parameter regimes nanoflare heating can also induce electron-ion non-equilibrium if the heating timescale is shorter than the electron-ion equilibration timescale. Interactions between electrons and ions in a fully-ionized hydrogen plasma like the solar corona are governed by binary Coulomb collisions. Thus, the equilbration timescale is $\tau_{ei}=1/\nu_{ei}$, where $\nu_{ei}$ is the collision frequency and is given by,
	\begin{equation}
		\label{eq:col_freq}
		\nu_{ei} = \frac{16\sqrt{\pi}}{3}\frac{e^4}{m_em_i}\left(\frac{2k_BT_e}{m_e}\right)^{-3/2}n\ln{\Lambda},
	\end{equation}
	where $T_e$ is the electron temperature, $m_e,m_i$ are the electron and ion masses respectively and $\ln{\Lambda}$ is the Coulomb logarithm \citep[see both Eq. 2.5e and Section 3 of][]{braginskii_transport_1965}. For $n\sim10^9$ cm$^{-3}$ and $T_e\sim10^{7}$ K, parameters typical of nanoflare heating, $\tau_{ei}\approx800$ s. Thus, any heating that occurs on a timescale less than 800 s, such as a nanoflare with a duration of $\tau\le100$ s, will result in electron-ion non-equilibrium. While chromospheric evaporation during and after the nanoflare will increase $n$ and thus decrease $\nu_{ei}$, we argue that during the early heating phase, $\tau_{ei}\gg\tau$, with 800 s being an upper bound on $\tau_{ei}$.
	\par While it is often assumed that the electrons are the recipients of the prescribed coronal heating function, ion heating in the solar corona should not be discounted since the exact mechanism behind coronal heating is still unknown. For example, ions may be heated via ion-cyclotron wave resonances \citep{markovskii_intermittent_2004} or magnetic reconnection \citep{ono_ion_1996,drake_onset_2014}. To address this possibility and include effects due to electron-ion non-equilibrium, we have applied the EBTEL analysis outlined in \citet{klimchuk_highly_2008} to the two-fluid hydrodynamic equations in the form given in the appendix of \citet{bradshaw_influence_2013}. Such an approach allows us to efficiently model a two-component impulsively-heated coronal plasma, and will be used extensively in \citetalias{barnes_inference_2016-1}.
	\par The two-fluid EBTEL equations are derived fully in \autoref{appendix_two_fluid} and are,
	\begin{align}
		\frac{d}{dt}\bar{p}_e &=\,\frac{\gamma - 1}{L}[\psi_{TR} - (\mathcal{R}_{TR} + \mathcal{R}_C)] + \nonumber \\ & k_B\bar{n}\nu_{ei}(\bar{T}_i-\bar{T}_e) + (\gamma-1)\bar{Q}_{e},\label{eq:press_e_0d_2fl} \\[0.5em]
		\frac{d}{dt}\bar{p}_i &=\,-\frac{\gamma - 1}{L}\psi_{TR} + k_B\bar{n}\nu_{ei}(\bar{T}_e-\bar{T}_i) + \nonumber \\ &(\gamma-1)\bar{Q}_{i},\label{eq:press_i_0d_2fl} \\[0.5em]
		\frac{d}{dt}\bar{n} &=\,\frac{c_2(\gamma-1)}{c_3\gamma Lk_B\bar{T}_e}(\psi_{TR} - F_{ce,0}-\mathcal{R}_{TR}).	\label{eq:mass_0d_2fl}
	\end{align}
	This set of equations is closed by the equations of state $p_e=k_BnT_e$ and $p_i=k_BnT_i$. While the notation above is largely self-evident, we draw attention to the additional term $\psi_{TR}$ which originates in the need to maintain charge and current neutrality and is defined by \autoref{eq:psi_TR}. Additionally, in both the single- and two-fluid versions of EBTEL used here, we have implemented an adaptive time-stepping routine to ensure that we are correctly resolving the thermal conduction timescale.
	\subsection{Ionization Non-equilibrium}
	\label{subsec:nei_theory}
	\par Ionization non-equilibrium has long been known to be an issue in the interpretation of data from the impulsive phase of flares, and more recently it has been discussed in the context of nanoflares \citep{bradshaw_explosive_2006,reale_nonequilibrium_2008}. The main issue is that when a tenuous plasma is heated rapidly, it takes a certain time to reach ionization equilibrium so that the ionization states present do not reflect the actual (electron) temperature, assuming that the heating occurs mainly to electrons (see \autoref{subsec:two_fluid_theory} and \autoref{subsec:two_fluid_res}) rather than the heavier ions such as Fe that contribute to the observed radiation. If the heating is sustained, then eventually ionization equilibrium will be reached, and this may occur in moderate to large flares. However, for nanoflares that may last for anywhere between a few seconds and a few minutes, a different scenario arises in which on termination of heating, rapid conductive cooling arises, so that the high ionization states may never be attained.
	\par \citet{bradshaw_explosive_2006}, \citet{reale_nonequilibrium_2008} and \citet{bradshaw_numerical_2009} have all addressed this point using slightly different approaches, but with similar conclusions, namely that short nanoflares in a low-density plasma are unlikely to be detectable. We now develop this work further to assess how the results in the first parts of \autoref{sec:results} are altered. We follow these authors and calculate an ``effective temperature'' ($T_{eff}$) as a proxy for the deviation from ionization equilibrium. This involves taking a time-series of $T$ and $n$ (e.g. from EBTEL) and using the numerical code\footnote{The numerical code used here has been made freely available by the author and is available at \url{https://github.com/rice-solar-physics/IonPopSolver.}} described in \citet{bradshaw_numerical_2009} to calculate the fractional ionization of as many states of various elements as needed, and in turn this calculates $T_{eff}$, a temperature that would be measured based on the actual ionization states. We primarily consider Fe between Fe IX and Fe XXVII, though Ca has also been calculated as a check on these results.
	\par The feature that will prove of great relevance in our results is that despite the different nanoflare durations, $T_{eff}$ does not exceed 10 MK. There is also an ``overshoot'' of $T_{eff}$ when it reaches its maximum value: this is saying that collisions are still not strong enough for the adjustment of the ionization state to be instantaneous.
	\section{Results}
	\label{sec:results}
	\par We now show a series of simulations of a single nanoflare with our zero-dimensional single- and two-fluid hydrodynamic EBTEL models, and the HYDRAD code. \citetalias{barnes_inference_2016-1} discusses long trains of multiple nanoflares of varying frequency in multiple loops. All results were processed using the IPython ecosystem \citep{perez_ipython:_2007} and the NumPy scientific computing package \citep{van_der_walt_numpy_2011}. All plots were produced using the matplotlib graphics environment \citep{hunter_matplotlib:_2007}.
	\par An important output of all these models is the coronal emission measure. In EBTEL the emission measure for the entire coronal part of the loop is calculated using the familiar expression $\mathrm{EM}=n^2(2L)$, where $L$ is the loop half-length. We consider a temperature range of $4.0\le\log{T}\le8.5$ with bin sizes of $\Delta\log{T}=0.01$. At each time $t_i$, the coronal temperature range $[T_0,T_a]$ is calculated from $\bar{T}$ ($\bar{T}_e$ for the two-fluid model). For each bin that falls within $[T_0,T_a]$, $\bar{n}_i^2(2L)$ is added to that bin, where $\bar{n}_i$ is the spatially-averaged number density at $t_i$. The emission measure in each bin is then averaged over the entire simulation period. When measured observationally, $\mathrm{EM}(T)$ is a line-of-sight quantity. Assuming an aspect ratio (i.e. ratio of loop length to loop width) of 10, we apply a correction factor 1/10 to all calculated $\mathrm{EM}$ curves. The emission measure from HYDRAD is calculated using quantities averaged over the upper 80\% of the loop which corresponds to the coronal portion of the loop.
	\subsection{Single-fluid Parameter Variations}
	\label{subsec:sf_par_var}
	\begin{figure*}
		\centering
		\begin{minipage}{0.49\textwidth}
			\subfigure{%
			\includegraphics[width=\columnwidth]{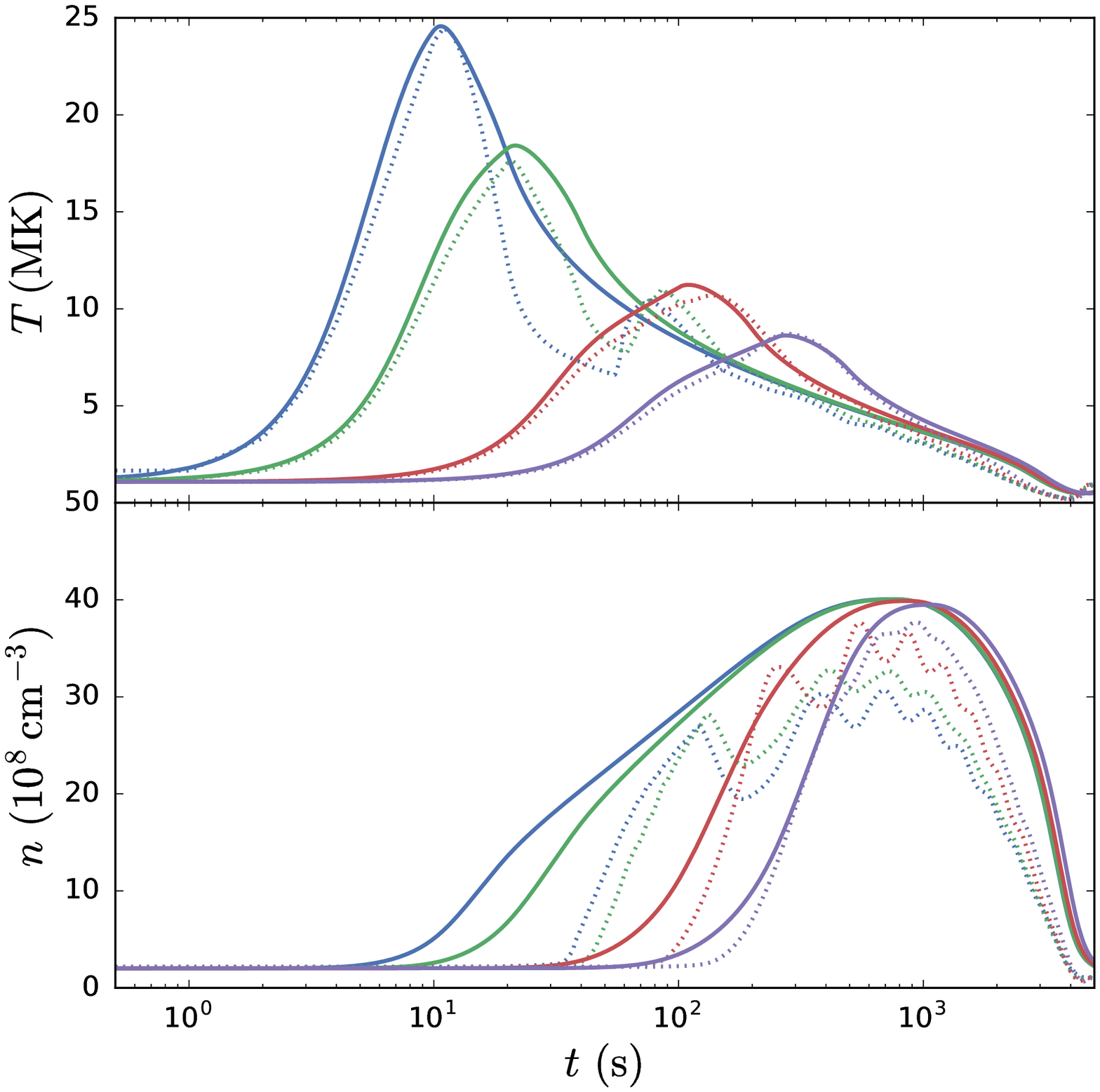}
			\label{fig:sf_T_panel1}}
		\end{minipage}
		\begin{minipage}{0.49\textwidth}
			\subfigure{%
			\includegraphics[width=\columnwidth]{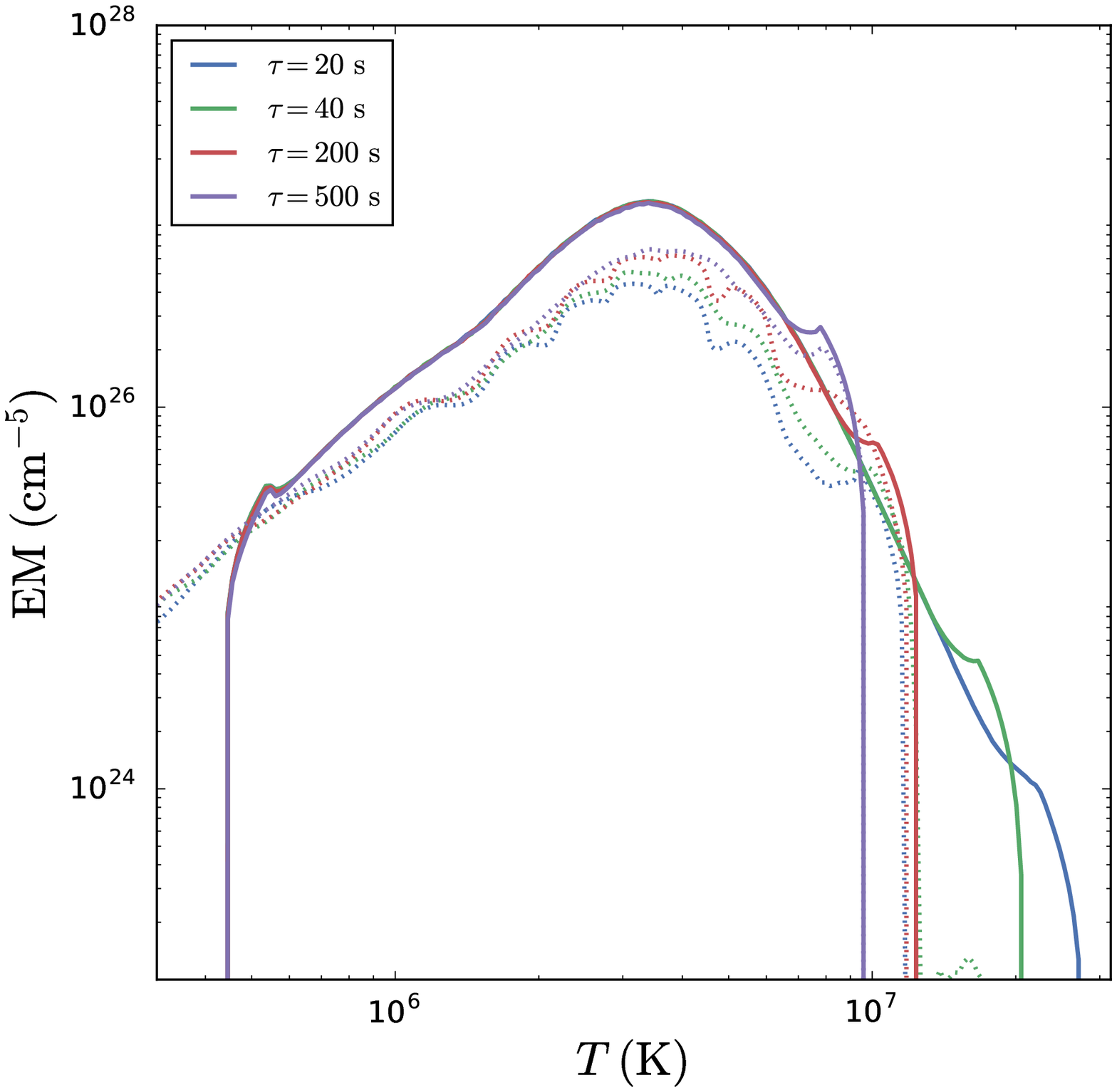}
			\label{fig:sf_em_panel3}}
		\end{minipage}
		\caption{\textbf{Left:} Temperature (upper panel) and density (lower panel) profiles for a loop with $2L=80$ Mm. Each heating profile is triangular in shape with a steady background heating of $H_{bg}=3.5\times10^{-5}$ ergs cm$^{-3}$ s$^{-1}$. The duration of the heating pulse is varied according to $\tau=20,40,200,500$ s, with each value of $\tau$ indicated by a different color, as shown in the right panel. The total energy injected into the loop is fixed at $10$ ergs cm$^{-3}$. Note that time is shown on a log scale to emphasize the behavior of the heating phase. \textbf{Right:} Corresponding $\mathrm{EM}(T)$ for each pulse duration $\tau$. The relevant parameters and associated colors are shown in the legend. $\mathrm{EM}(T)$ is calculated according to the procedure outlined in the beginning of \autoref{sec:results}. In all panels, the solid (dotted) lines show the corresponding EBTEL (HYDRAD) results (see \autoref{subsubsec:hydrad_comparison_sf}).}
		\label{fig:sf_tnem}
	\end{figure*}
	\subsubsection{Varying Pulse Duration}
	\label{subsubsec:pulse_res}
		\par In the first set of results we assume the plasma behaves as a single fluid, use a flux limiter of $f=1$, and ignore ionization non-equilibrium. The solid curves in \autoref{fig:sf_tnem} show average temperature (upper left panel) and density (lower left panel) as a function of time for a single nanoflare in a loop with $2L = 80$ Mm where the EBTEL approach is used.  The heating function takes the form of a triangular pulse for four different pulse durations, $\tau=$ 20, 40, 200, and 500 s, as indicated by the legend in the right panel. The peak heating rate is varied such that the total energy input is 10 ergs cm$^{-3}$ for all cases. These parameters correspond roughly to bright AR core loops \citep{warren_systematic_2012}. In order to ensure that the temperature and density do not become negative, a small background heating of magnitude $H_{bg}=3.5\times10^{-5}$ ergs cm$^{-3}$ s$^{-1}$ is enforced at all times. It can be seen that shorter pulses give higher temperatures, as expected. Furthermore, in this early heating phase, one would expect the maximum temperature to scale roughly as $H_0^{2/7}$ (where $H_0$ is the peak heating rate); this is approximately what is found. On the other hand, the different pulse durations give approximately the same maximum density, with the shortest pulse reaching its peak value roughly 200 s before the longest.
		\par The solid lines in the right panel of \autoref{fig:sf_tnem} show the corresponding EBTEL emission measure distributions, $\mathrm{EM}(T)$. The temperature of maximum emission ($T_m$) and the peak emission measure ($\mathrm{EM}(T_m)$)are the same in all cases and are consistent with those found in the studies of AR core loops \citep[e.g.][]{warren_systematic_2012}. While shorter pulses lead to higher initial temperatures, the shape of the emission measure below $T_m$ is independent of the properties of the heating pulse, indicating that this part of the emission measure distribution cannot provide information about the actual nanoflare duration or intensity. All cases show evidence of the heating phase, namely the bump on $\mathrm{EM}(T)$ at $\log{(T)} =$ 6.85, 7, 7.2 and 7.3. Below these bumps to just above $T = T_m$, $\mathrm{EM}(T)$ scales as $T^{-5}-T^{-5.5}$ for all cases, again indicating that information about the heating process is lost at these temperatures. However, detection of emission above $T_m$ in a single structure would still be evidence for nanoflare heating, though of undetermined duration.

		\par For integration over the lifetime of unresolved structures lying transverse to the line of sight, one can write down an expression $\mathrm{EM}(T) \sim n^2\tau_{cool}(n, T)$ which simply states that what matters for determining $\mathrm{EM}(T)$ is how long the plasma spends at any given temperature \citep[e.g.][]{cargill_implications_1994,cargill_nanoflare_2004}. For an analytic solution for the cooling, one can formally define $\tau_{cool}(n, T) = (T/(dT/dt))$. In the absence of a formal solution, order of magnitude scalings can be used: the difference with analytic solutions being a numerical factor. To obtain an expression $\mathrm{EM}(T)\propto T^{-b}$, one needs to provide a relation between $T$ and $n$. For conductive cooling of the corona, one can write $\tau_{cool} \sim nL^2T^{-5/2}$, giving $\mathrm{EM} \sim n^3L^2T^{-5/2}$. In determining the relationship between $T$ and $n$, two limits are those of constant density and constant pressure. The former gives static conductive cooling \citep[e.g.][]{antiochos_influence_1976} and the latter evaporative cooling with constant thermal energy \citep[e.g.][]{antiochos_evaporative_1978}, which then lead to $b = 5/2$ and $11/2$ respectively. Fitting the EBTEL $\mathrm{EM}(T)$ results for $\tau\le200$ s (see right panel of \autoref{fig:sf_tnem}) to $T^{-b}$ on $10^{6.8}<T<10^{7}$ K yields $b\sim4.5-5$ which are more consistent with the latter.
	\subsubsection{HYDRAD Comparison}
	\label{subsubsec:hydrad_comparison_sf}
		\par We now compare EBTEL and HYDRAD results for the different values of $\tau$. The dotted lines in all three panels of \autoref{fig:sf_tnem} show the corresponding HYDRAD results, where averaging is over the upper 80\% of the loop. The background heating in the two codes has been adjusted to ensure that EBTEL and HYDRAD start with the the same initial density since the initial temperature rise will depend on the assumed background density.
		\par There is good agreement between the HYDRAD and EBTEL results for $\tau\ge200$ s with the well-documented result that EBTEL gives somewhat higher density maxima than HYDRAD \citep[see][]{cargill_enthalpy-based_2012}. For $\tau=20,40$ s, while the peak temperatures are at a level of agreement consistent with previous work \citep{cargill_enthalpy-based_2012}, there are notable differences in the initial temperature decay from the maximum in the upper left panel of \autoref{fig:sf_tnem} due to the difference in the initial density response.
	\par It can be seen that the EBTEL density begins to rise almost immediately following the onset of heating, while there is a lag in the HYDRAD density. This is due to a delay in the upflow of material from the TR because a finite time is required to get material moving up the loop, an effect absent from 0D models. The slower density rise seen with HYDRAD leads to the faster conductive cooling. Another feature of the short pulses is the very spiky density profile as a function of time. This is a well-known effect, particularly in flare simulations, and is due to pairs of oppositely-directed flows colliding at the loop top, and subsequently bouncing back and forth.
	\par As a result of this discrepancy in the density behavior, while the emission measure calculated from the EBTEL model ``sees'' temperatures well in excess of 10 MK for short pulses, in the HYDRAD model this will not be the case. This is evident from the short pulses in the right panel of \autoref{fig:sf_tnem}: the emission above 10 MK predicted by EBTEL is not present in the HYDRAD runs, the emission cutting off just above $10^7$ K. For the longer pulses, EBTEL still shows emission at higher temperatures, but the difference with HYDRAD is evident now over a smaller temperature range. Also, the characteristic bumps on the emission measure seen with EBTEL are largely eliminated in the HYDRAD runs.
	\par This regime of short heating pulses was not considered in our earlier work using EBTEL, and the associated comparisons with field-aligned hydrodynamic codes \citep{klimchuk_highly_2008,cargill_enthalpy-based_2012}, where pulses of order 200 s or greater were considered. Other workers have used short pulses with EBTEL, albeit much less intense \citep{tajfirouze_euv_2016,tajfirouze_time-resolved_2016}. Clearly the more gentle the heating profile used, the slower the rise in the EBTEL density, leading to results closer to those found using HYDRAD. Thus it appears that caution is warranted in the use of approximate models for short, intense heating pulses. This restriction only applies to the high temperature regime: as can be seen from \autoref{fig:sf_tnem}, the emission measure profiles below $10^{6.8}$ are not affected. Nonetheless, the absence of emission near 10 MK for short pulses constitutes one of many obstacles to quantifying any hot plasma component due to nanoflares.
	\subsubsection{Heat Flux Limiter}
	\label{subsubsec:hf_res}
	\begin{figure}
		\centering
		\includegraphics[width=\columnwidth]{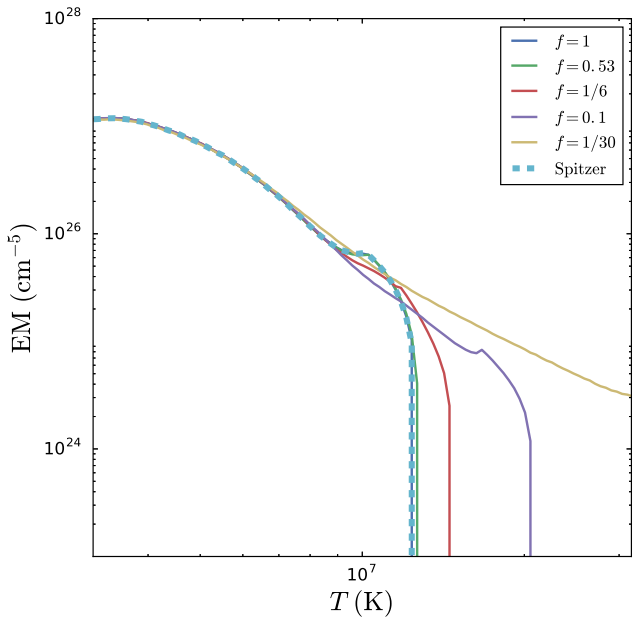}
		\caption{$\mathrm{EM}(T)$ calculated from the single-fluid EBTEL model when only pure Spitzer conduction is used (turquoise, dashed) and when a flux limiter is imposed according to \autoref{subsec:hf_theory}. In the free-streaming limit, \textbf{five different values of $f$ are considered (see legend)}. The pulse duration is $\tau=200$ s. All other parameters are the same as those discussed in \autoref{subsubsec:pulse_res}. Note that here we only show $\mathrm{EM}(T)$ for $T>T_m$ \textbf{as the cool side of $\mathrm{EM}(T)$ is unaffected by our choice of $f$.}}
		\label{fig:sf_hf_em}
	\end{figure}
	\par \autoref{fig:sf_hf_em} shows the effect of using a flux limiter versus Spitzer conduction on the emission measure distribution. Five different values of $f$ are shown: 1 \citep[blue,][consistent with HYDRAD]{bradshaw_influence_2013}, 0.53 \citep[green,][]{karpen_nonlocal_1987}, $1/6$ \citep[red,][]{patsourakos_coronal_2005}, 0.1 \citep[purple,][]{luciani_nonlocal_1983}, and $1/30$ (yellow). The pulse duration is 200 s and only the EBTEL results are shown. Note that for this pulse length, the HYDRAD results are expected to be similar.
	\par As expected, inclusion of a limiter extends $\mathrm{EM}(T)$ to higher temperatures, though this is only notable above 10 MK. As the temperature falls to this value, evaporative upflows have increased the coronal density so that the Spitzer description is recovered. Above 10 MK flux limiting gradually becomes important, albeit with a small emission measure. Using $f=0.53,1$ yield $\mathrm{EM}(T)$ that are not discernibly different from that produced by pure Spitzer conduction while $f=1/6,0.1$ extend $\mathrm{EM}(T)$ to significantly hotter temperatures. $f=1/30$, the most extreme flux limiter, yields an emission well above $10^{7.5}$ K. Note that for all cases, $\mathrm{EM}(T)$ converges to the same value for $T\le10$ MK.
	\par For flux-limited thermal conduction, $\tau_{cool} \sim LT^{-1/2}$ so that the parameter $b$ lies between 1/2 and 5/2, depending on the assumption about $n$. For $f = 1/30$, $b = 5/2$ is found in \autoref{fig:sf_hf_em} by fitting $\mathrm{EM}(T)$ to $T^{-b}$ on $10^7\le T\le10^{7.5}$ K. Since the free-streaming limit slows conduction cooling relative to that given by Spitzer, the plasma will spend more time at any given temperature, leading to smaller values of $b$. Similar conclusions hold for other conduction models \citep[e.g. the non-local model discussed in the coronal context by][]{karpen_nonlocal_1987,west_lifetime_2008} since they all inhibit conduction. While limiting of conduction is often regarded as an important process in coronal cooling, these results suggest that for nanoflare heating it may not be that important unless extreme values of the limiting parameter are used.
	\subsection{Two-fluid Effects}
	\label{subsec:two_fluid_res}
	\subsubsection{Electron Heating}
	\label{subsubsec:electron_heating}
	\begin{figure*}
		\centering
		\begin{minipage}{0.49\textwidth}
			\subfigure{%
			\includegraphics[width=\columnwidth]{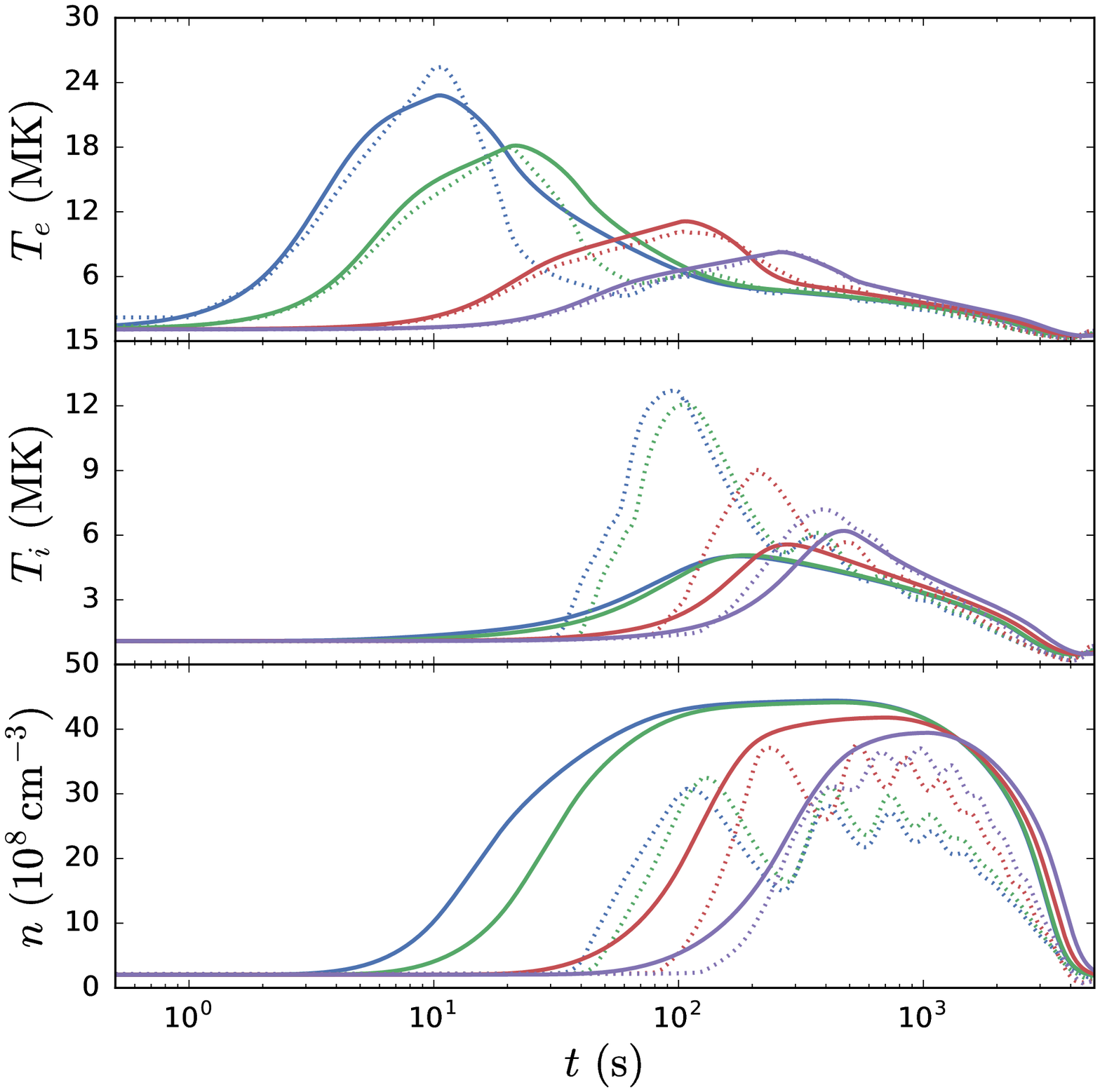}
			\label{fig:tfe_T_panel1}}
		\end{minipage}
		\begin{minipage}{0.49\textwidth}
			\subfigure{%
			\includegraphics[width=\columnwidth]{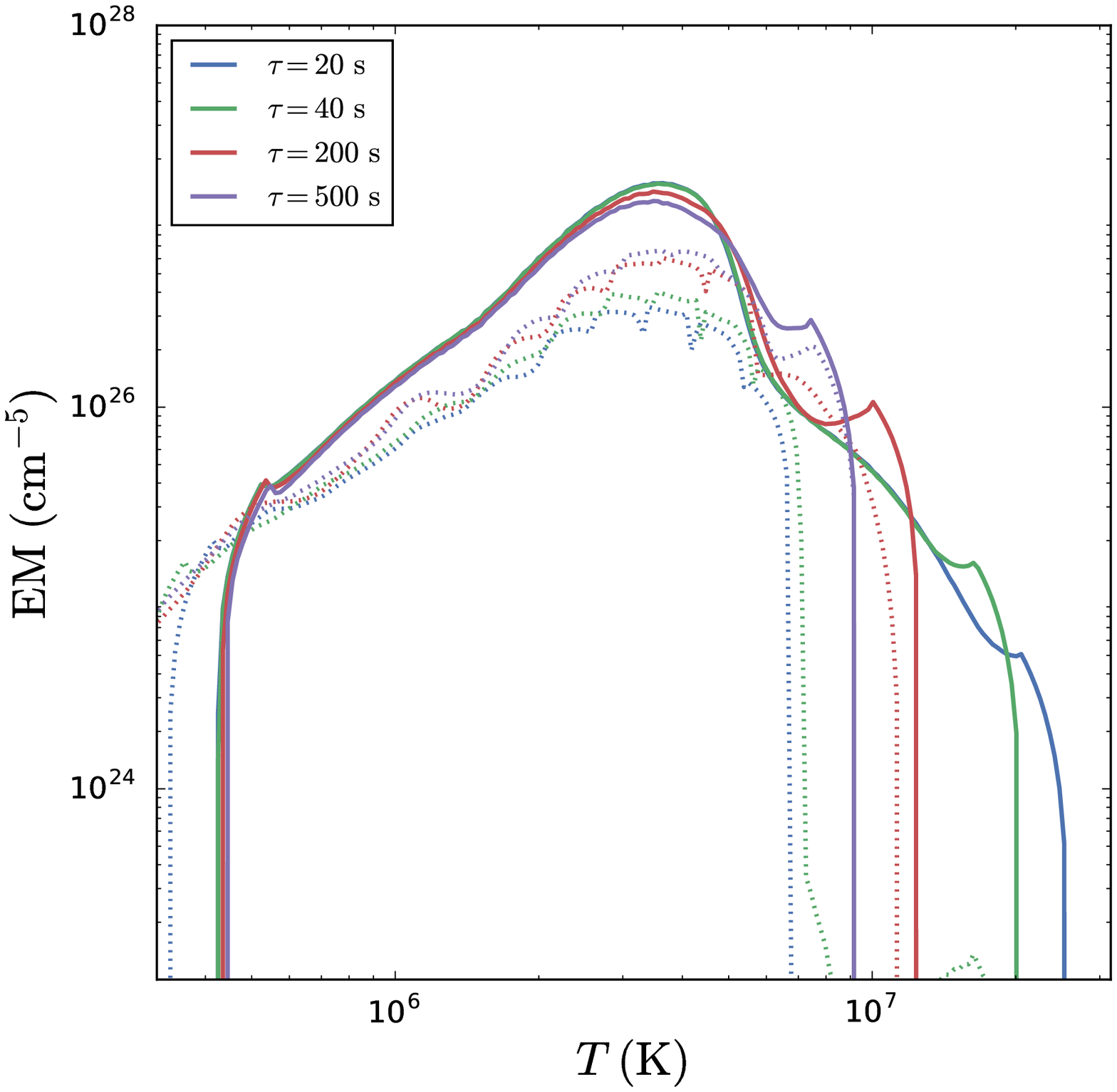}
			\label{fig:tfe_em_panel3}}
		\end{minipage}
		\caption{Two-fluid EBTEL simulations for $\tau=20,40,200,500$ s in which only the electrons are heated. \textbf{Left}: Electron temperature (upper panel), ion temperature (middle panel), and density (lower panel). \textbf{Right:} Corresponding $\mathrm{EM}(T)$ calculated according to \autoref{sec:results}. The pulse durations and associated colors for all panels are shown on the right. All parameters are the same as those discussed in \autoref{subsubsec:pulse_res}. In all panels, the solid (dotted) lines show the corresponding EBTEL (HYDRAD) results.}
		\label{fig:tfe_tnem}
	\end{figure*}
		\par We now use our two-fluid model to consider the role of separate electron or ion heating, focusing on cases when only the electrons or ions are heated in order to highlight the essential difference between the two scenarios. Intermediate cases of energy distribution will be considered in subsequent papers. The solid lines in the left panels of \autoref{fig:tfe_tnem} show the electron temperature (upper panel), ion temperature (middle panel) and density (lower panel) as a function of time from the two-fluid EBTEL model for $\tau=20,40,200,500$ s for electron heating. The dotted lines show the corresponding HYDRAD results and are discussed in \autoref{subsubsec:hydrad_comparison_tf} The electrons now cool by a combination of thermal conduction and temperature equilibration, the latter becoming significant at 150 (450) s for short (long) pulses. The ions thus heat rather slowly, reaching a peak temperature of 5 MK, which overshoots the electron temperature at that time. The ions then cool via ion thermal conduction and equilibration, with $T_e \approx T_i$ after typically a few hundred seconds.
	\par The solid lines in the right panel of \autoref{fig:tfe_tnem} show the resulting $\mathrm{EM}(T)$. In the case of electron heating and $\tau<500$ s, the emission measure slope over the temperature interval $\log{T_M}<\log{T}<6.8$ is considerably steeper compared to the single-fluid case. Recall that in the single-fluid case we assume that conduction is the only relevant cooling mechanism prior to the onset of radiative cooling such that under the assumption of constant pressure, $\mathrm{EM}\propto T^{-11/2}$ (see \autoref{subsec:sf_par_var}). When we allow for electron-ion non-equilibrium and heat only the electrons, both of these assumptions break down. Following the onset of conductive cooling, $T_e\gg T_i$, but the loop has now begun to fill. The equilibration term plays the part of a cooling term so long as $T_e>T_i$ and is the dominant cooling mechanism for several hundred seconds in between the peak electron temperature and the peak density (see \autoref{fig:psi_tr_compare}). Thus, our expression for $\tau_{cool}$ should include some contribution from the equilibration term in this temperature regime.
	\begin{figure}
		\centering
		\includegraphics[width=\columnwidth]{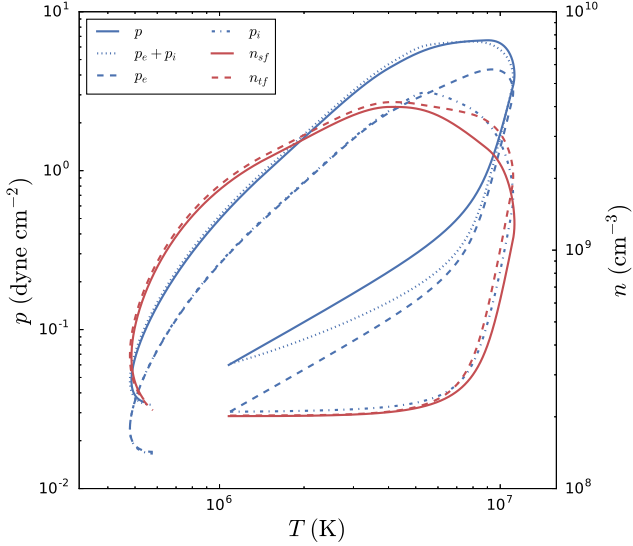}
		\caption{Pressure (left axis, blue lines) and density (right axis, red lines) as a function of temperature for the $\tau=200$ s case. All parameters are the same as those discussed in \autoref{subsubsec:pulse_res}. The single-fluid pressure $p$ and density $n$ are denoted by the solid blue and red lines, respectively. The two-fluid total pressure, $p_e+p_i$, electron pressure, $p_e$, and ion pressure, $p_i$, are denoted by the dotted, dashed, and dot-dashed blue lines respectively. The two-fluid density is represented by the dashed red line. Pressure, density, and temperature are all shown on a log scale.}
		\label{fig:pnt_state_space}
	\end{figure}
	\par\autoref{fig:pnt_state_space} shows pressure (blue lines) and density (red lines) as a function of temperature for the $\tau=200$ s case; both the single-fluid case and the case where only the electrons are heated are shown. While $p_e+p_i$ (blue dotted line), the total pressure, like the single-fluid pressure $p$ (blue solid line) is constant over the interval $10^{6.65}<T<10^{6.8}$, the electron pressure, $p_e$ (blue dashed line) is not, meaning $n\propto T_e^{-1}$ is not a valid scaling law in the two-fluid, electron-heating case. Comparing the two-fluid density (dashed red line) and the single-fluid density (solid red line) easily confirms this. To derive a emission measure slope for the case in which only the electrons are heated, these effects must be accounted for in the $\mathrm{EM}(T)\sim n^2\tau_{cool}(n,T)$ scaling. Thus, while a power-law $b$ may be calculated by fitting the hot part of the $\mathrm{EM}(T)$ to $T^{-b}$, it is difficult to gain any physical insight from such a fit using the scaling discussed in \autoref{subsubsec:pulse_res}.
	\subsubsection{Ion Heating}
	\label{subsubsec:ion_heating}
	\begin{figure*}
		\centering
		\begin{minipage}{0.49\textwidth}
			\subfigure{%
			\includegraphics[width=\columnwidth]{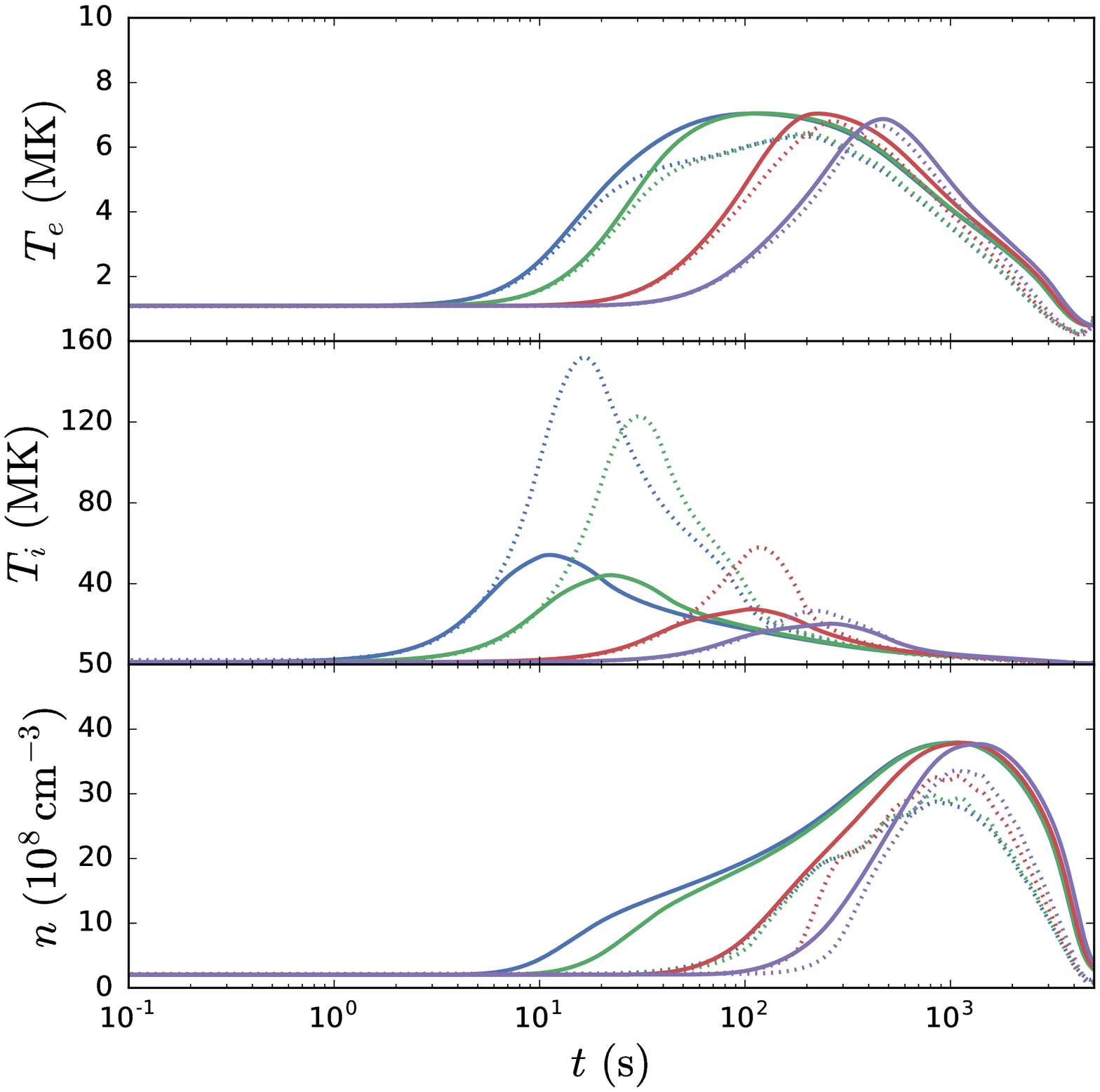}
			\label{fig:tfi_T_panel1}}
		\end{minipage}
		\begin{minipage}{0.49\textwidth}
			\subfigure{%
			\includegraphics[width=\columnwidth]{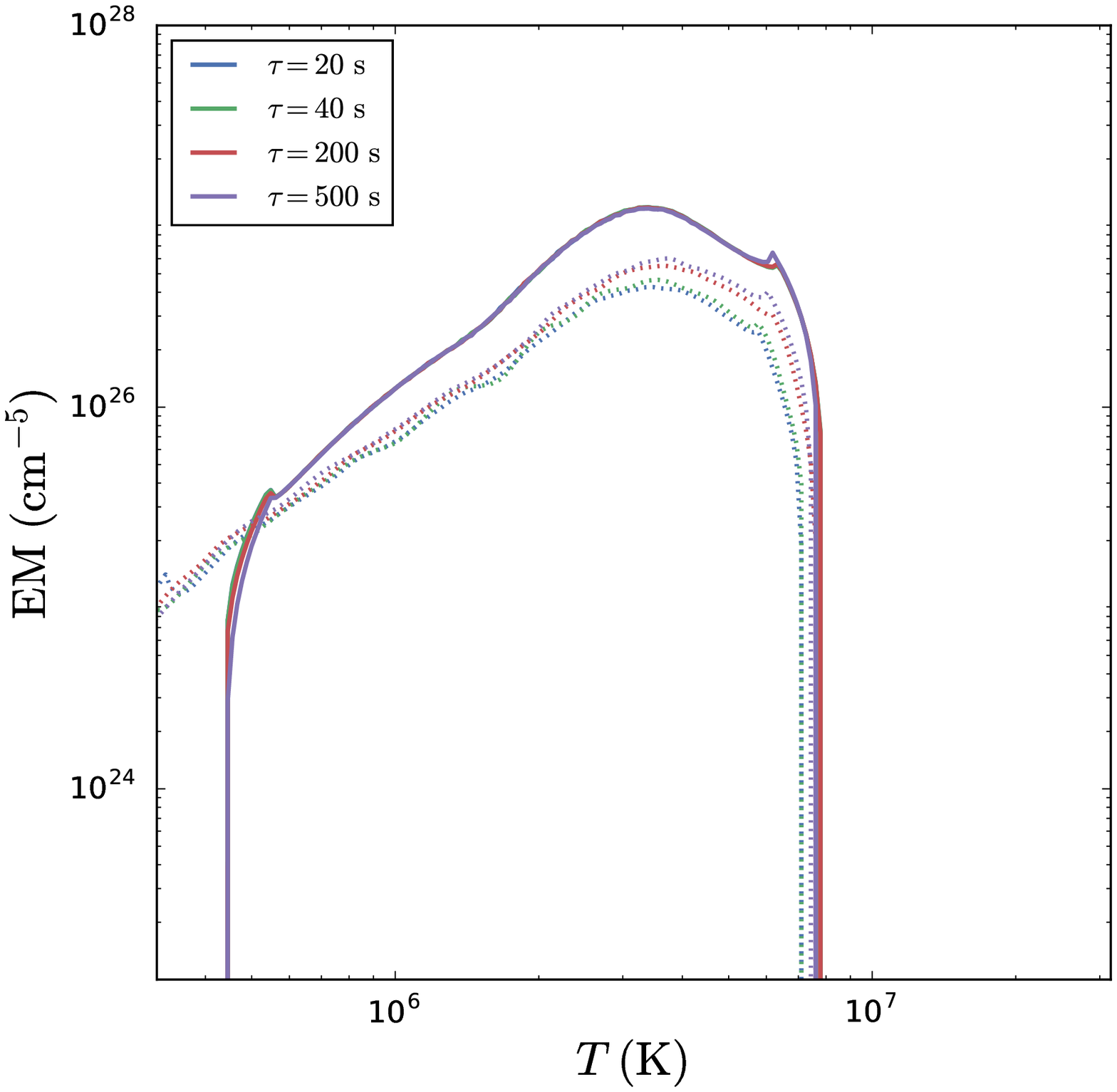}
			\label{fig:tfi_em_panel3}}
		\end{minipage}
		\caption{Two-fluid EBTEL simulations for $\tau=20,40,200,500$ s in which only the ions are heated. \textbf{Left}: Electron temperature (upper panel), ion temperature (middle panel), and density (lower panel). \textbf{Right:} Corresponding $\mathrm{EM}(T)$ calculated according to \autoref{sec:results}. The pulse durations and associated colors for all panels are shown on the right. All parameters are the same as those discussed in \autoref{subsubsec:pulse_res}. In all panels, the solid (dotted) lines show the corresponding EBTEL (HYDRAD) results.}
		\label{fig:tfi_tnem}
	\end{figure*}
	\autoref{fig:tfi_tnem} shows the electron temperature (upper left panel), ion temperature (middle left panel), density (lower left panel) and the corresponding emission measure (right panel) for $\tau=20,40,200,500$ s when only the ions are heated. The solid lines show the two-fluid EBTEL results while the dotted lines show the corresponding HYDRAD results (see \autoref{subsubsec:hydrad_comparison_tf}). Ion heating leads to significantly higher temperatures due to the relative weakness of ion thermal conduction, consistent with the expected enhancement of $(\kappa_{0,e}/\kappa_{0,i})^{2/7}$. The hot ions cool by a combination of weak ion thermal conduction and temperature equilibration. However, because the Coulomb coupling timescale during the early heating phase (when $T_i\gg T_e$ and the density is low) is much larger than the ion thermal conduction timescale, by the time the electrons can ``see'' the ions, they have cooled far below their peak temperature. The peak electron temperature in all cases lies below 10 MK. Because $\mathrm{EM}(T)$ is constructed from the electron temperature, the emission measure never sees $T\ge10^7$ K, with $\mathrm{EM}(T)$ being truncated sharply near $10^{6.9}$ K for all values of $\tau$.
	\par The reason for slower equilibration for ion heating can be seen by comparing the density plots in the lower left panels of \autoref{fig:tfe_tnem} and \autoref{fig:tfi_tnem}. These show that while the peak values of the density are similar for both heating mechanisms, the temporal behavior differs for ion heating with shorts pulses: for these cases, the density takes considerably longer to reach the maximum value. This can be attributed to the relative weakness of ion thermal conduction. Examination of \autoref{eq:0d_mass_sub} and \autoref{eq:psi_TR} shows that an upward enthalpy flux can only be effective for ion heating once temperature equilibration has become significant and an electron heat flux is established. In turn, once the upflow begins, the coronal density increases, making equilibration more effective. Thus, once temperature equilibration starts to be effective, these processes combine to give a rapid increase in density, as shown.
	\par In the case where the heating pulse duration is long, $\tau=500$ s, the difference between the two-fluid and single-fluid emission measure distributions is diminished. Because the electrons are heated slowly, they do not have much time to evolve out of equilibrium with the ions. This in turn heavily dampens the Coulomb exchange term, allowing the two populations to evolve together as a single fluid.
	\subsubsection{HYDRAD Comparison}
	\label{subsubsec:hydrad_comparison_tf}
	\par The dotted lines in all panels of \autoref{fig:tfe_tnem} and \autoref{fig:tfi_tnem} show the corresponding HYDRAD results for both electron and ion heating, respectively. As in \autoref{subsubsec:hydrad_comparison_sf}, the averaging is done over the upper 80\% of the loop and the background heating has been adjusted appropriately. For $\tau\ge200$ s, we find acceptable agreement in $n$, $T_e$, and $\mathrm{EM}(T)$.
	\par For $\tau=20,40$ s, the upper and lower panels of \autoref{fig:tfe_tnem} show discrepancies in $T_e$ and $n$ similar to those discussed in \autoref{subsubsec:hydrad_comparison_sf}. The initial decay from the peak electron temperature is noticably different in the EBTEL runs compared to the corresponding HYDRAD runs, again due to the difference in the initial density response. The discrepancies in the density are exacerbated in the electron heating case (compared to the single-fluid case) since all of the energy is partitioned to the electrons, resulting in a stronger electron heat flux and a subsequently stronger upflow. The right panel of \autoref{fig:tfe_tnem} shows the effect of this premature rise in the density on $\mathrm{EM}(T)$ for these short pulses: while EBTEL predicts significant emission above 10 MK, the emission in the HYDRAD runs cuts off just below $10^{6.9}$ K.
	\par In the ion heating case, we find acceptable agreement in $T_e$ and $\mathrm{EM}(T)$ despite similar discrepancies in $n$ for the shortest heating pulses, $\tau=20,40$ s. Because no heat is supplied to the electrons directly, the electron heating timescale is set by the Coulomb collision frequency (see \autoref{eq:col_freq}), meaning energy is deposited to the electrons over a timescale much longer than 20 or 40 s. The resulting slow evolution of $T_e$ leads to subsequently weaker upflows. Because of the much more gentle rise in density, the electrons are not able to ``see'' the ions until they have cooled well below 10 MK (see \autoref{subsubsec:ion_heating}).
	\par In the middle panels on the left-hand side of \autoref{fig:tfe_tnem} and \autoref{fig:tfi_tnem}, the ion temperature in HYDRAD is greater than that of EBTEL by a factor of $\sim3-4$ in the late heating/early conductive cooling phase. These spikes in $T_i$ are due to steep velocity gradients that heat the ions through compressive heating and viscosity, two pieces of physics that are not included in EBTEL. Because ion thermal conduction is comparatively very weak, these sharp features in $T_i$ are not as efficiently smoothed out. While these differences in $T_i$ are more prominent when $\tau=20,40$ s, they still persist for $\tau\ge200$ s.
	\subsection{Ionization Non-equilibrium}
	\label{subsec:nei_res}
	\begin{figure*}
		\centering
		\includegraphics[width=2\columnwidth]{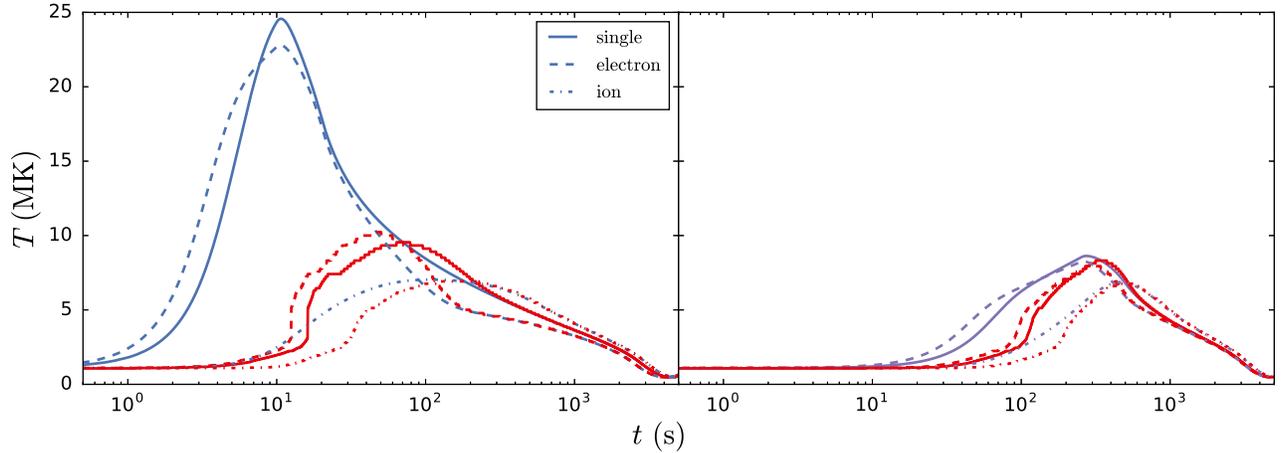}
		\caption{$T_{eff}$ (red) for pulse durations of 20 s (left panel) and 500 s (right panel) for the single-fluid case (solid) as well as the cases where only the electrons (dashed) or only the ions (dot-dashed) are heated. $T(t)$ profiles (i.e. assuming ionization equilibrium) for $\tau=20$ s (blue lines) and $\tau=500$ s (purple lines) for all three heating scenarios are repeated here for comparison purposes.}
		\label{fig:stf_Teff_compare}
	\end{figure*}
	\begin{figure*}
		\centering
		\includegraphics[width=2\columnwidth]{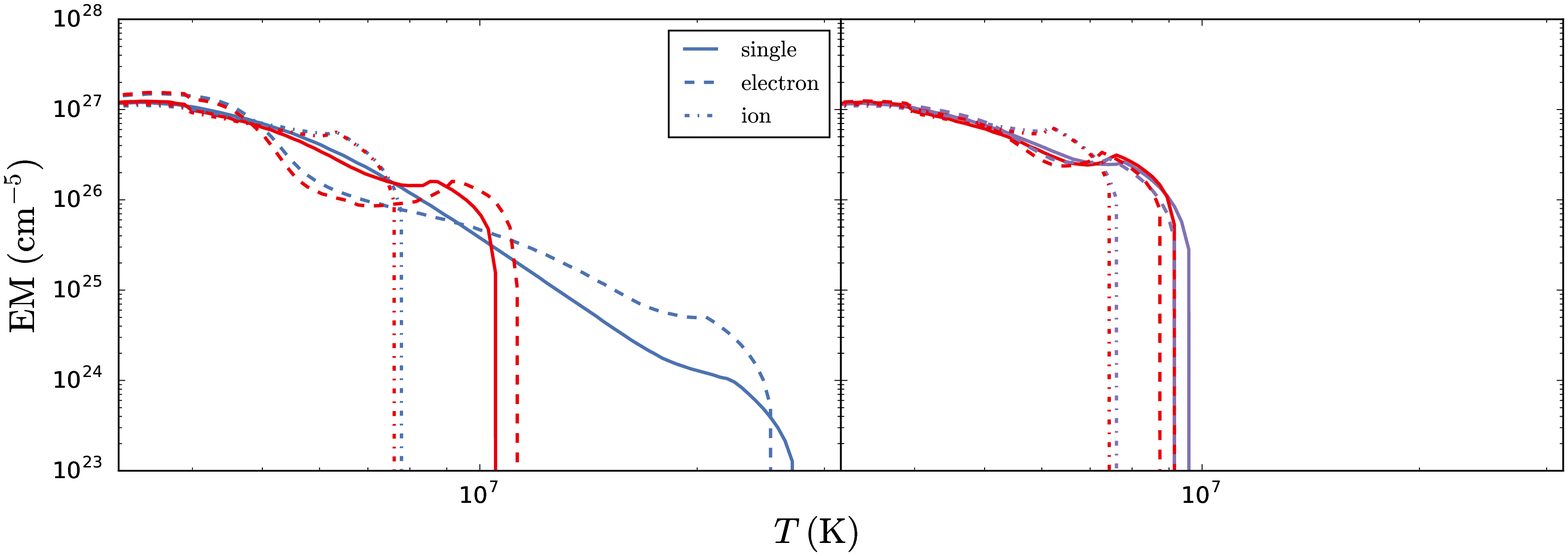}
		\caption{$\mathrm{EM}(T_{eff})$ (red) for pulse durations of 20 s (left panel) and 500 s (right panel) for the single-fluid (solid), electron heating (dashed), and ion heating (dot-dashed) cases. $\mathrm{EM}(T)$ (i.e. assuming ionization equilibrium) for $\tau=20$ s (blue lines) and $\tau=500$ s (purple lines) for all three heating scenarios are repeated here for comparison purposes. Note that in both panels we only show $\mathrm{EM}(T)$ for $\log{T}>\log{T_M}$.}
		\label{fig:stf_emeff_compare}
	\end{figure*}
	\par The final set of results includes our approximate treatment of non-equilibrium ionization, again using the EBTEL approach. The red curves in the left (right) panel of \autoref{fig:stf_Teff_compare} show $T_{eff}$ for $\tau=20\,(500)$ s for the single-fluid, electron heating, and ion heating cases. For comparison, equivalent results for $T$ (single-fluid) and $T_e$ (two-fluid) that assume ionization equilibrium are shown. For all cases, $T_{eff}$ never rises above 10 MK for the short pulse and 8 MK for the long pulse. Thus, for the short pulse, because a sufficiently long time is required to ionize the plasma, the hottest electron temperatures are never likely to be detectable. For the longer pulse, the slow heating gives the ionization states the opportunity to ``catch up''; thus $T_{eff}$ is a reasonable reflection of the actual plasma state.
	\par The red curves in \autoref{fig:stf_emeff_compare} show the corresponding $\mathrm{EM}(T_{eff})$. The effect of ionization non-equilibrium is to truncate  $\mathrm{EM}$ around or below 10 MK. The bump on the distribution characteristic of the heating phase is also relocated to lower temperatures. This confirms the earlier comment that, at least for short pulses, the hot electron plasma above 10 MK is undetectable. While the heating signature is shifted to smaller values of $T_{eff}$, one has no way of knowing the duration of the pulse that generates it. Thus it seems as if the temperature range $T_m<T<10$ MK is the optimal one for searching for this hot component as well as direct signatures of the heating. However, it is difficult to ``map'' what would be seen in such a state of ionization non-equilibrium back to the real system.
	\section{Discussion}
	\label{sec:discussion}
	\par This paper has begun to address signatures of the so-called ``hot'' plasma component in the non-flaring corona, especially ARs, that is perceived as providing essential evidence for the existence of nanoflares. In this first paper in a series, we have used zero-dimensional and field-aligned single- and two-fluid modeling to examine the possible signatures of a single nanoflare occuring in a low-density plasma. This corresponds to the simplest case of so-called ``low frequency'' (LF) nanoflares, where a coronal loop is heated by many events with the same energy and with a time between events longer than the characteristic cooling time such that the plasma is allowed cool significantly before being reenergized.
	\par When an approximate single-fluid model assuming ionization equilibrium is used, the expected signatures of conductive cooling appear in the distribution of plasma as a function of temperature, as described by the emission measure. In particular, short nanoflares with duration under 100 sec should have a significant plasma component well above 10 MK, and for longer duration events, significant plasma between the temperature of the maximum emission measure and 10 MK. However, inclusion of several pieces of additional physics modifies this result considerably, in each case making it much less likely that any plasma that is above 10 MK can be detected.
	\par For short nanoflares, the time taken for conductively-heated chromospheric plasma to move into the coronal part of a loop is sufficiently long that the initial hot coronal plasma cools rapidly, contributing little to the emission measure such that, once the coronal density has increased, its temperature is below 10 MK. This effect is less important for long duration nanoflares. Consideration of separate electron and ion heating shows that, while electron heating leads to similar results to the single fluid case, ion heating results in no emission measure at 10 MK due to the principal electron heating mechanism being a relatively slow collisional process. Finally, relaxing the assumption of ionization equilibrium leads to a truncation of the emission measure below 10 MK, since the time needed to create highly ionized states such as Fe XXI is longer than any relevant cooling time. In all cases the hot plasma, while still in the corona, is effectively ``dark''. In addition, characteristic structures in the emission measure profile that are a signature of the heating itself in simple models are all but eliminated.
	\par These results suggest that while showing that such a ``hot'' plasma should exist in principle may not be difficult, characterizing the heating process from its observed properties may be a lot harder. Of course we have limited ourselves to the LF nanoflares here, and we showed \citep{cargill_active_2014} that the intermediate frequency nanoflare regime does have significant differences, in large part due to the range of densities that the nanoflares occur in. This will be addressed fully, along with other parameter variations, in \citetalias{barnes_inference_2016-1}, though it is difficult to see how a component hotter than 10 MK can be resurrected. Note though that the results of \citet{caspi_new_2015} pose a challenge for our scenario unless an undetected microflare or small flare occurred during the observations.
	\par The observational aspects of this work will be addressed more fully in \citetalias{barnes_inference_2016-1}. However, one can conclude (i) present day observations do not seem capable of making quantitative statements about the ``hot'' component, though they are highly suggestive of its existence and (ii) future measurements should be concentrated in the temperature regime $10^{6.6}$ – $10^7$ K rather than at higher temperatures. The MaGIXS instrument, due to fly in 2017, is well positioned to do this.
\acknowledgments
WTB was provided travel support to the Coronal Loops Workshop
VII held in Cambridge, UK, July 21-23, 2015, at which a preiliminary version of this work was presented, by NSF award number 1536094. We thank the anonymous referee whose comments helped to improve the final draft of this paper.

	\appendix
	\section{Modifications to $c_1$ during the Conductive Cooling Phase}
	\label{appendix_c1_corrections}
	\par In Section 3 of \citet{cargill_enthalpy-based_2012} we assumed that the parameter $c_1$ decreased from its equilibrium value at the time of maximum density, to that commensurate with radiative/enthalpy cooling as the loop drained. This was defined in terms of the ratio $n/n_{eq}$, where $n_{eq}$ was the loop density that would exist for the calculated temperature were the loop to be in static equilibrium \citep[Equation 17 of][]{cargill_enthalpy-based_2012}. In this radiative phase, $n > n_{eq}$. On the other hand, when $n < n_{eq}$, we assumed $c_1$ took on its equilibrium value, $c_{1,eq}$. Defining $\Delta\equiv(n_{\mathrm{EBTEL}} - n_{\mathrm{HYDRAD}})/n_{\mathrm{HYDRAD}}$, this gave $\Delta\lesssim0.2$, acceptable errors in the EBTEL value of $n$, as shown in the figures in \citet{cargill_enthalpy-based_2012}, in particular for the mild nanoflares we considered.
	\par It is now clear that a modified description of $c_1$ for $n < n_{eq}$ is needed for many of the examples discussed in the present paper. Specifically, for intense heating events, the coronal density calculated by the version of EBTEL in \citet{cargill_enthalpy-based_2012} is unacceptably high when compared to results from the HYDRAD code. Quantitatively, we find $\Delta\gtrsim0.3$ at $n_{max}$. While this may seem to be reasonable for an aproximate model, the high EBTEL density is a systematic feature, and requires further investigation.
	\par Examination of the HYDRAD results shows that EBTEL significantly underestimates the TR radiative losses during the heating and conductive cooling phases. At this time, the loop is under-dense \citep[e.g.][]{cargill_nanoflare_2004}, so that an excess of the conducted energy goes into evaporating TR material. We have modified $c_1$ as follows for $n < n_{eq}$,
	\begin{equation}
		c_1 = \frac{2c_{1,eq} + c_{1,cond}((n_{eq}/n)^2-1)}{1+(n_{eq}/n)^2},
		\label{eq:c1_mod}
	\end{equation}
	as a direct analogy to Eq. 18 of \citet{cargill_enthalpy-based_2012}. In the early phases of an event, $n \ll n_{eq}$, so that $c_1 \approx c_{1,cond}$. When $n = n_{eq}$, $c_1 = c_{1,eq}$. After some experimentation, we have settled on a choice of $c_{1,cond} = 6$ since that gives reasonable agreement between EBTEL and HYDRAD. There is no impact on the solution for $n > n_{eq}$.
	\par\autoref{tab:table_c1_compare} shows a set of runs we have carried out to compare the results from HYDRAD and EBTEL with $c_1=c_{1,eq}=2$ (fifth column) and with $c_1$ given by \autoref{eq:c1_mod} (sixth column), when $n<n_{eq}$. We find that using the modification in \autoref{eq:c1_mod} gives, for the more intense heating cases with $\tau\ge200$ s, $\Delta\sim0.1$ at $n_{max}$. For the more gentle heating profiles of \citet{cargill_enthalpy-based_2012} and \citet{bradshaw_influence_2013} (i.e. rows 3, 4, 6, and 8 of \autoref{tab:table_c1_compare}), we continue to find $\Delta\lesssim0.2$, confirming that the modification proposed here is applicable to a wide range of heating scenarios. For short, intense pulses like the $\tau=20,40$ s cases addressed in this paper, we still find $\Delta>0.2$. We have addressed the limitations of such cases in \autoref{subsubsec:hydrad_comparison_sf}.
	\begin{deluxetable}{cccccc}
\tablecaption{Comparison between HYDRAD and EBTEL with $c_1=2$ and $c_1$ given by \autoref{eq:c1_mod}, for $n<n_{eq}$.
The first three columns show the full loop length, heating pulse duration, and maximum heating rate.
The last three columns show $n_{max}$ for the three models.
Only $n_{max}$ is shown as $T_{max}$ is relatively insensitive to the value of $c_1$. 
The first two rows correspond to the $\tau=200,500$ s cases considered in this paper. 
The next four rows are the four cases shown in Table 2 of \citet{cargill_enthalpy-based_2012}. 
The last two rows are cases 6 and 11 from Table 1 of \citet{bradshaw_influence_2013}.
}
\tablehead{\colhead{$2L$} & \colhead{$\tau$} & \colhead{$H_0$} & \colhead{$n_{max}$, HYDRAD} & \colhead{$n_{max}$, EBTEL} & \colhead{$n_{max}$, EBTEL (Eq. \ref{eq:c1_mod})}\\ \colhead{(Mm)} & \colhead{(s)} & \colhead{(erg cm$^{-3}$ s$^{-1}$)} & \colhead{($10^8$ cm$^{-3}$)} & \colhead{($10^8$ cm$^{-3}$)} & \colhead{($10^8$ cm$^{-3}$)}}
\startdata
80 & 200 & 0.1 & 37.6 & 44.2 & 39.6 \\
80 & 500 & 0.04 & 37.7 & 44.1 & 39.3 \\
150 & 500 & 0.0015 & 3.7 & 3.8 & 3.4 \\
50 & 200 & 0.01 & 10.7 & 11.3 & 10.1 \\
50 & 200 & 2 & 339.0 & 391.8 & 351.0 \\
50 & 200 & 0.01 & 15.5 & 16.3 & 14.3 \\
40 & 600 & 0.8 & 350.0 & 452.9 & 391.0 \\
160 & 600 & 0.005 & 10.0 & 10.2 & 9.1
\enddata
\label{tab:table_c1_compare}
\end{deluxetable}

	\par\autoref{eq:c1_mod} is motivated by simplicity while including the essential physics. Alternative, more complex determinations of $c_1$ have been considered, but involve limitations on how EBTEL can be used both now and in the future.
	\section{Derivation of the Two-fluid EBTEL Equations}
	\label{appendix_two_fluid}
	\par The two-fluid field-aligned hydrodynamic mass and energy equations, as given by \citet{bradshaw_influence_2013}, are:
	\begin{align}
		\frac{\partial\rho}{\partial t} &= -\frac{\partial(\rho v)}{\partial s} \label{eq:1dmass} \\[0.5em]
		\frac{\partial E_e}{\partial t} + \frac{\partial}{\partial s} \lbrack(E_e+p_e)v\rbrack &= v\frac{\partial p_e}{\partial s} - \frac{\partial F_{ce}}{\partial s} + \frac{1}{\gamma - 1}k_Bn\nu_{ei}(T_i-T_e) -n^2\Lambda(T_e)+Q_{e} , \label{eq:1denergy_e} \\[0.5em]
		\frac{\partial E_i}{\partial t} + \frac{\partial }{\partial s}\lbrack(E_i+p_i)v\rbrack &= -v\frac{\partial p_e}{\partial s} - \frac{\partial F_{ci}}{\partial s} + \frac{1}{\gamma - 1}k_Bn\nu_{ei}(T_e-T_i) + \frac{\partial}{\partial s}\left(\frac{4}{3}\mu_iv\frac{\partial v}{\partial s}\right) +\rho v g_{\parallel} + Q_{i},\label{eq:1denergy_i}
	\end{align}
	where,
	\begin{align}
		E_e =& \frac{p_e}{\gamma - 1} \label{eq:ee_closure}, \\[0.5em]
		E_i =& \frac{p_i}{\gamma - 1} + \frac{\rho v^2}{2}, \label{eq:ei_closure}
	\end{align}
	and we assume closure through the ideal gas law, $p_e=k_BnT_e,\,p_i=k_BnT_i$. Note that we have assumed quasi-neutrality such that $n_e=n_i=n$ and $v_e=v_i=v$. It then follows that $\rho=m_en_e+m_in_i\approx m_in$.
	\par Note the right-hand side of \autoref{eq:1denergy_e} and \autoref{eq:1denergy_i}: the first term represents the energy loss or gain as the fluids move through the electric field that maintains quasi-neutrality, given by $E=-(1/ne)\partial p_e/\partial s$; the third term models the exhange of energy between the electron and ion populations via binary Coulomb collisions and is attributed to \citet{braginskii_transport_1965}. Though the expression presented here differs by a factor of 2 compared to that of \citeauthor{braginskii_transport_1965}, we maintain that the electron-ion equilibration time is not significantly changed by this relatively small numerical factor.
	\par Plugging in these expressions for $E_e$ and $E_i$ and using the assumptions of sub-sonic flows ($v<C_s$) and loops shorter than a gravitational scale height ($L<150$ Mm) as outlined in \citet{klimchuk_highly_2008}, the two-fluid field-aligned hydrodynamic energy equations can be written,
	\begin{align}
		\frac{1}{\gamma - 1}\frac{\partial p_e}{\partial t} + \frac{\gamma}{\gamma - 1}\frac{\partial}{\partial s}(p_ev) &= v\frac{\partial p_e}{\partial s} - \frac{\partial F_{ce}}{\partial s} + \frac{1}{\gamma - 1}k_Bn\nu_{ei}(T_i-T_e) -n^2\Lambda(T_e)+Q_{e}, \label{eq:1denergy_e_simp} \\[0.5em]
		\frac{1}{\gamma - 1}\frac{\partial p_i}{\partial t} + \frac{\gamma}{\gamma - 1}\frac{\partial }{\partial s}(p_iv)&= -v\frac{\partial p_e}{\partial s} - \frac{\partial F_{ci}}{\partial s} + \frac{1}{\gamma - 1}k_Bn\nu_{ei}(T_e-T_i) + Q_{i}. \label{eq:1denergy_i_simp}
	\end{align}
	Notice that we have dropped the ion viscous and gravitational terms from  \autoref{eq:1denergy_i} as well as the kinetic energy term from \autoref{eq:ei_closure}. $Q_{e}$ and $Q_{i}$ represent the electron and ion heating terms, respectively. $F_{ce}$ and $F_{ci}$ are the electron and ion heat flux terms, respectively. In the case of Spitzer conduction, $\kappa_{0,e}=7.8\times10^{-7}$ and $\kappa_{0,i}=3.2\times10^{-8}$.
	\par The analysis now follows that of \citet{klimchuk_highly_2008} and \citet{cargill_enthalpy-based_2012}. Assuming symmetry about the loop apex, we integrate \autoref{eq:1denergy_e_simp} and \autoref{eq:1denergy_i_simp} over the coronal loop half-length $L$,
	\begin{align}
		\frac{L}{\gamma - 1}\frac{d \bar{p}_e}{dt} &= \frac{\gamma}{\gamma - 1}(p_ev)_0 + F_{ce,0} + \psi_C - \mathcal{R}_C + L\bar{Q}_{e},\label{eq:1denergy_e_C} \\[0.5em]
		\frac{L}{\gamma - 1}\frac{d \bar{p}_i}{dt} &= \frac{\gamma}{\gamma - 1}(p_iv)_0 + F_{ci,0} - \psi_C + L\bar{Q}_{i},\label{eq:1denergy_i_C}
	\end{align}
	where we have assumed the enthalpy flux and heat flux go to zero at the loop apex, $R_C=\int_C\mathrm{d}s\,n^2\Lambda(T_e)$ and,
	\begin{equation}
		\psi_C=\int_C\mathrm{d}s\,v\frac{\partial p_e}{\partial s} + \int_C\mathrm{d}s\,\frac{k_B}{\gamma - 1}n\nu_{ei}(T_i - T_e).
	\end{equation}
	\par Similarly, integrating over the TR portion of the loop of thickness $\ell$, we obtain,
	\begin{align}
		\frac{\gamma}{\gamma - 1}(p_ev)_0 &= - F_{ce,0} + \psi_{TR} - \mathcal{R}_{TR}, \label{eq:1denergy_e_TR} \\[0.5em]
		\frac{\gamma}{\gamma - 1}(p_iv)_0 &=  - F_{ci,0} - \psi_{TR}, \label{eq:1denergy_i_TR}
	\end{align}
	where several terms are neglected because $\ell\ll L$ \citep{klimchuk_highly_2008}. Additionally, we have assumed that the enthalpy flux and heat flux go to zero at the top of the chromosphere, $R_{TR}=\int_{TR}\mathrm{d}s\,n^2\Lambda(T_e)$ and
	\begin{equation}
		\psi_{TR}=\int_{TR}\mathrm{d}s\,v\frac{\partial p_e}{\partial s} + \int_{TR}\mathrm{d}s\,\frac{k_B}{\gamma - 1}n\nu_{ei}(T_i - T_e).
	\end{equation}
	The second term in this expression is usually small, but is retained for completeness.  Plugging \autoref{eq:1denergy_e_TR} (\autoref{eq:1denergy_i_TR}) into \autoref{eq:1denergy_e_C} (\autoref{eq:1denergy_i_C}),
	\begin{align}
		\frac{L}{\gamma - 1}\frac{d\bar{p}_e}{dt} =& \psi_{TR} + \psi_C -(\mathcal{R}_C + \mathcal{R}_{TR}) + L\bar{Q}_{e},\label{eq:0d_press_e_sub} \\[0.5em]
		\frac{L}{\gamma - 1}\frac{d\bar{p}_i}{dt} =& -(\psi_{C} + \psi_{TR}) +  L\bar{Q}_{i}.\label{eq:0d_press_i_sub}
	\end{align}
	Note that adding \autoref{eq:0d_press_e_sub} and \autoref{eq:0d_press_i_sub} gives the correct single-fluid EBTEL model (i.e. \autoref{eq:energy_0d}).
	\par As in the single-fluid case, we find that the spatially-integrated coronal density evolution is described by,
	\begin{equation}
		L\frac{d\bar{n}}{dt} = (nv)_0.
	\end{equation}
	Using \autoref{eq:1denergy_e_TR} and the equation of state for $p_e$, the above equation can be written as
	\begin{align}
		(nv)_0 =& \frac{(p_ev)_0}{k_BT_{e,0}} = \frac{c_2(\gamma - 1)}{c_3\gamma k_B\bar{T}_e}(-F_{ce,0} - \mathcal{R}_{TR} + \psi_{TR}),\\
		L\frac{d\bar{n}}{dt} =& \frac{c_2(\gamma - 1)}{c_3\gamma k_B\bar{T}_e}(-F_{ce,0} - \mathcal{R}_{TR} + \psi_{TR}).\label{eq:0d_mass_sub}
	\end{align}
	\par To obtain \autoref{eq:press_e_0d_2fl}, \autoref{eq:press_i_0d_2fl}, and \autoref{eq:mass_0d_2fl}, we need to find expressions for $\psi_C$ and $\psi_{TR}$. Recall that $\psi_C$ and $\psi_{TR}$ are comprised of terms associated with the quasi-neutral electric field and temperature equilibration. The integral of the former can be considered as the gain or loss of energy associated with plasma motion through the net electric potential. Consider the first integral in the definition of $\psi_C$. Using integration by parts,
	\begin{equation}
		\int_C\mathrm{d}s\,v\frac{\partial p_e}{\partial s} = (p_ev)\Big|^{``a"}_{``0"} - \int_C\mathrm{d}v\,p_e = -(p_ev)_0 - \int_C\mathrm{d}v\,p_e\approx -(p_ev)_0 -\bar{p}_e\int_C\mathrm{d}v = -(p_ev)_0 + \bar{p}_ev_0 \approx 0.
	\end{equation}
	Thus, we can express $\psi_C$ as
	\begin{equation}
		\psi_C\approx\frac{k_BL}{\gamma -1}\bar{n}\nu_{ei}(\bar{T}_i - \bar{T}_e),
		\label{eq:psi_C}
	\end{equation}
	where $\nu_{ei}=\nu_{ei}(\bar{T}_e,\bar{n})$. To find an expression for $\psi_{TR}$, we first note that, using the equation of state for both the electrons and the ions and the quasi-neutrality condition ($n_e=n_i$),
	\begin{equation}
		\frac{p_ev}{p_iv} = \frac{T_e}{T_i}.
	\end{equation}
	Evaluating this expression at the TR/corona interface (denoted by ``0''), plugging in \autoref{eq:1denergy_e_TR} and \autoref{eq:1denergy_i_TR},
	\begin{equation}
		\frac{- F_{ce,0} + \psi_{TR} - \mathcal{R}_{TR}}{- F_{ci,0} - \psi_{TR}} = \xi,
	\end{equation}
	where $\xi\equiv T_{e,0}/T_{i,0}$. Solving for $\psi_{TR}$, we find,
	\begin{equation}
		\psi_{TR} = \frac{1}{1+\xi}(F_{ce,0} + \mathcal{R}_{TR} - \xi F_{ci,0}).
		\label{eq:psi_TR}
	\end{equation}
	\par Plugging \autoref{eq:psi_C} and \autoref{eq:psi_TR} into \autoref{eq:0d_press_e_sub}, \autoref{eq:0d_press_i_sub}, and \autoref{eq:0d_mass_sub} gives us our set of two-fluid EBTEL equations as given in \autoref{eq:press_e_0d_2fl}, \autoref{eq:press_i_0d_2fl}, and \autoref{eq:mass_0d_2fl}. The prescription for $c_1$, $c_2$, and $c_3$ is the same as the single-fluid version of EBTEL. As discussed in \citet{cargill_enthalpy-based_2012}, these play little role in the early heating phase when two-fluid effects are important.
	\begin{figure}[t]
		\centering
		\includegraphics[width=0.5\columnwidth]{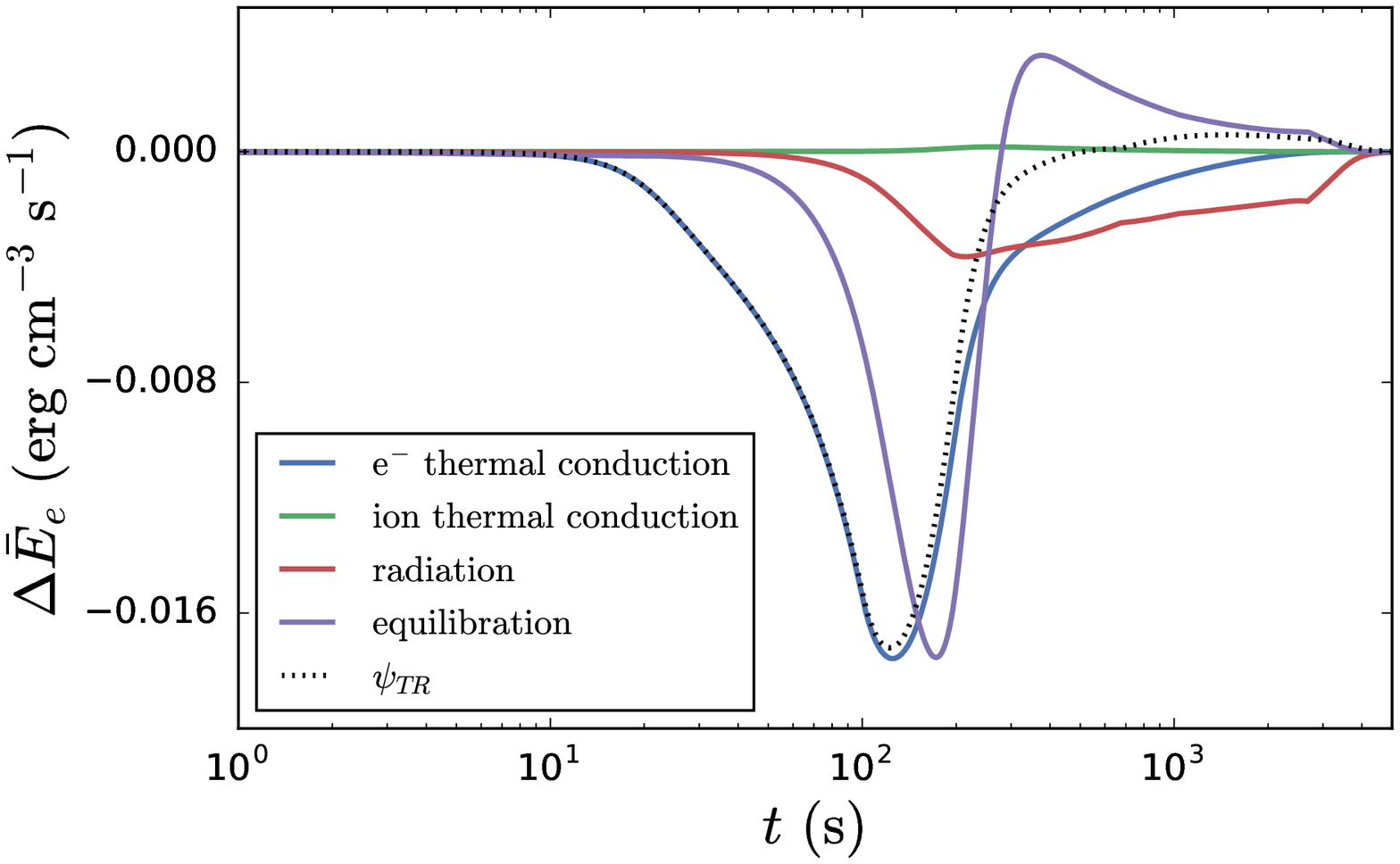}
		\caption{Energy loss and gain mechanisms arising from a nanoflare with $\tau = 200$ s and electron heating only. The various curves correspond to the terms in the EBTEL two-fluid electron energy equation (\autoref{eq:press_e_0d_2fl_breakdown}). Electron and ion thermal conduction, radiation, binary Coulomb interactions, and $\psi_{TR}$ are shown. The loop parameters are as in \autoref{sec:results}.}
		\label{fig:psi_tr_compare}
	\end{figure}
	\par Plugging \autoref{eq:psi_TR} into \autoref{eq:press_e_0d_2fl}, the electron energy evolution equation can be written,
	\begin{equation}
		\frac{1}{\gamma -1}\frac{d\bar{p}_e}{dt} = \frac{1}{L(1+\xi)}F_{ce,0} - \frac{\xi}{L(1+\xi)}F_{ci,0} - \frac{\xi(c_1+1) + 1}{L(1+\xi)}\mathcal{R}_C + \frac{k_B}{\gamma-1}\bar{n}\nu_{ei}(\bar{T}_i-\bar{T}_e) + \bar{Q}_e,
		\label{eq:press_e_0d_2fl_breakdown}
	\end{equation}
	where the first two terms on the right-hand side represent the contributions from electron and ion thermal conduction, the third term represents losses from radiation, and the last two terms are as before. \autoref{fig:psi_tr_compare} shows the contribution of each term, with the exception of the heating term, $\bar{Q}_e$. As expected, (electron) thermal conduction dominates during the early heating and cooling phase and losses from radiation takeover in the late draining and cooling stage. Between these two phases, energy exchange between the two species is important to the evolution of the electron energy. $\psi_{TR}$, indicated by the black dotted line, is included to show its importance in the formation of the two-fluid EBTEL equations.
	\bibliography{paper.bib}
	\bibliographystyle{aasjournal}
\end{document}